\documentclass[letterpaper]{article}
\usepackage{graphicx}
\usepackage{amssymb}
\usepackage{amsmath}
\usepackage{amsfonts,enumerate}
\begin{document}

\title{Differentiated QoS with Modified C/I Based Scheduling Algorithm}
\author{A.Divya and Sanjay Singh\thanks { Sanjay Singh is with the Department of Information and Communication Technology, Manipal Institute of Technology, Manipal University, Manipal-576104, India, E-mail: sanjay.singh@manipal.edu}}

\maketitle

\begin{abstract}
Second-generation (2G) digital cellular systems constitute the majority of cellular communication deployed today.  A variety of services of 2G systems has increased significantly and this will continue to grow even further in the emerging third-generation (3G) systems. Universal Mobile Telecommunication System (UMTS) is a third-generation mobile communications system which uses the Wide-Band Code Division Multiple Access (WCDMA) technique to support a wide variety of services, like speech, video telephony, Internet browsing, etc. These services require a wide range of Quality of Service (QoS) requirements. QoS is an important issue as the number of multimedia services increases day by day. Differentiated QoS methods allow the differentiation of users based on their priority levels and channel conditions so that the network can allocate the bandwidth for a particular request based on the QoS requirements. These requirements are controlled by Radio Resource Management (RRM) mechanisms. In this paper we have proposed two RRM algorithms which are modification to the existing scheduling algorithms. One is Prioritized C/I scheduling, which takes the priorities into consideration, and this algorithm serves the user with highest priority. Other algorithm is Modified Inverse C/I scheduling, which takes channel conditions into consideration and serves the users in degraded conditions, thereby improving QoS. The performance evaluation of two algorithms is done with EURANE extensions for NS-2. Simulation results shows the improvement in QoS for the users who are at equidistance from Base Station (BS) but requesting for different services by implementing Prioritized C/I scheduling and also for the users who are in degraded conditions by implementing Modified Inverse C/I scheduling when compared to Max C/I and Inverse C/I scheduling algorithm respectively.

\end{abstract}

\section{Introduction}
\label{sec1}
Cellular systems have started initially with voice communication. Several new data applications such as short-message service (SMS), electronic mail, WAP (wireless application protocol) for the web micro-browsing, and multimedia message service (MMS) have been gradually included \cite{2}. Inclusion of more data-centric services in new mobile networks and the requirement of better support for those services need more and better network interconnection options as compared to the 2G cellular systems. As a result, the enhanced 2G networks and their successor 3G systems came into  existence. With the increasing number of multimedia services in 3G systems the Quality of service (QoS) support is not straight forward, because all the services are to be carried out using band limited wireless channel. Hence there is a need of some QoS management techniques to improve the service quality for different type of services. This can be achieved through differentiation of mobile users based on his/her priority level and also based on channel conditions. Differentiated QoS refers to adjusting the allotted bandwidth for a particular request (e.g., voice call, Web browsing) to achieve acceptable service levels pre-determined by a user. 
\par
A. Jamalipour et al \cite{1} observed that with multimedia services and more network interconnections, support of QoS in cellular networks would not be direct anymore and also observed the effect of window size on transmission rate of the TCP sender with and without delay spikes and found that with the presence of the delay spikes the average throughput is decreasing; increasing the initial window size, and thus the performance in cellular networks with delay spikes is degraded. Mario Vranješ et al \cite{3} observed that large amount of radio resources are allocated to the users who are closer to the base stations. The UMTS performance is observed in two environments namely pedestrian and rural and the comparison of Max C/I based scheduling algorithm is done in terms of delay and throughput. The results showed that, as a consequence of such scheduling mechanism, data transfer performance for closer users are better than for farther users and also the users that are close to the BS, the user speed has a significant impact on data transfer performance, while for very far users the losses are quite high and thus there is no big difference whether the user moves slowly or fast.
\par
V. Siris et al \cite{4} have presented models for fair and efficient service differentiation in 3G mobile networks which involves allocating resources based on weights in uplink and downlink and investigated the effect of Signal to Interference Ratio (SIR) estimation errors, and discrete transmission rates on service differentiation. The major limitation of this work is that the traffic is not considered to be adaptive for both the transmission rate and also the signal quality. Riri Fitri Sari et al \cite{16} did comparison study on multimedia services in UMTS using HS-DSCH. The comparison is done by considering two modes in Radion Link Control (RLC) viz Acknowledged Mode (AM) and Unacknowledged Mode (UM) and found that RLC in AM is suitable for QoS classes namely web server and File Transfer Protocol (FTP) traffic.
\par
The 3G systems such as UMTS and HSDPA were not part of the mainstream NS-2 code. The implementation of UMTS and HSDPA functionalities in NS-2 is provided in \cite{7}.  Performance comparison of different RLC modes using different multimedia services for UMTS is given in \cite{7}. Masmoudi et al \cite{12} has proposed the radio resource and scheduling optimization in HSDPA based on UMTS networks, while the performance evaluation of an Enhanced UMTS network in a business city center environment is presented in \cite{17}.
\par
By considering all the proposed methods by different authors and concentrating on future enhancements of the work, this paper gives simulation results of an Enhanced UMTS mobile network which implements HSDPA and examines its behavior by measuring network performances (end-to-end delay) for mobile users who are located in different environment namely Pedestrians, Rural, Hilly, Urban, and Indoor. We have modified two existing C/I based scheduling algorithm. The scheduling is being done at the base station (BS). In one scheduling algorithm we consider the improvement to existing Max C/I scheduling, the modified algorithm is being called as Prioritized C/I scheduling. The Prioritized C/I scheduling algorithm implemented takes the Channel Quality Indicator (CQI) values and priorities as inputs and does the scheduling process. The channel used between the mobile station and Base station is High Speed Downlink Shared Channel (HS-DSCH) and Transmission Time Interval (TTI) considered is 2ms. The scheduler checks the queue for every 2ms interval, hence the average end to end delay is comparatively less when we use HS-DSCH instead of DSCH channel.
\par
Another algorithm is  Modified Inverse C/I scheduling which takes the CQI values as inputs and does the scheduling, it considers all the users as same (as in Round Robin) and gives preference to the user away from BS. Using both the algorithms we can say that the QoS has improved when delay constraint is considered. The performance of Prioritized C/I scheduling is found to be better than Max C/I in all the environments considered. 
\par
The rest of the paper is organized as follows. Section 2 describes the existing two C/I based scheduling algorithm and proposes their modified version in order to improve QoS. Section 3 discusses about the simulation results obtained and finally section 4 concludes this paper.
\section{Proposed Scheduling Techniques}
The present work is carried out in ns2.30. The Network Simulator (NS-2) is a discrete, event driven network simulator developed initially at UC Berkeley \cite{5} and later supported by the Virtual Inter Network Testbed (VINT) project \cite{vnit}. EURANE extensions were added to NS-2 which adds three nodes namely RNC (Radio Network Control), BS (Base Station), and UE (User Equipment). EURANE includes two main scheduling algorithms namely:
\begin{itemize} 
\item Max C/I Scheduling and 
\item Round Robin Scheduling.
\end{itemize}
Max C/I scheduling considers the channel conditions of the users and based on the CQI values it does the scheduling process for the mobile users who are equidistant from the BS. This algorithm serves the user with highest CQI value \cite{7}. The limitation here is, it does not consider the users with the priorities, i.e even when a high priority user is there in the network, he/she will not be served unless their CQI value is high.
\par
Our Proposed Algorithm modifies the Max C/I scheduling algorithm. In this algorithm we consider the users whose priorities are high and who are equidistance from BS requesting for different type of services. This algorithm serves the user with highest priority. If two users have same priorities then the scheduling is done based on CQI values. In this way when compared to Max C/I scheduling, by using our method the accessing delay for a high priority user is reduced, there by QoS will be improved. Filipe Leitao et al \cite{18} has proposed the Inverse C/I scheduling algorithm which has been implemented by tuning C/I scheduling. The drawback observed here is that all the users are randomly grouped into three classes namely Gold, Silver and Bronze by considering Gold has high priority and bronze has the least priority, independent of their channel conditions. This way we cannot identify which users request has come first to the BS.
\par
\begin{figure}[!t]
\includegraphics[width=12cm]{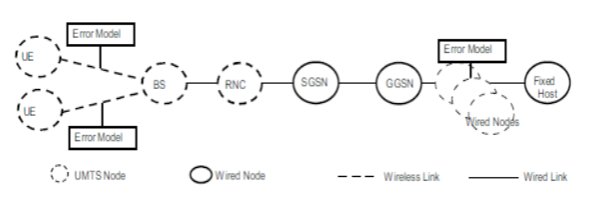}
\caption{NS-2 Simulation Model}
\label{fig:1}
\end{figure}
\begin{figure}[!t]
\includegraphics[width=12cm]{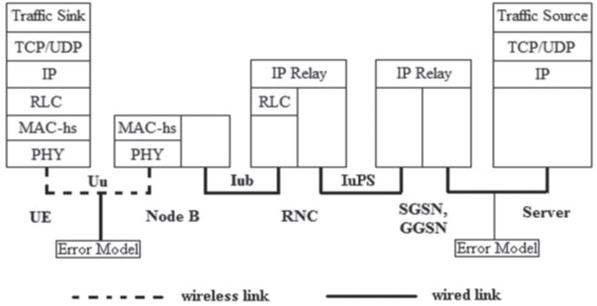}
\caption{HSDPA Protocol Architecture \cite{7}}
\label{fig:2}
\end{figure}

In this paper we also present Modified Inverse C/I scheduling which takes channel conditions into consideration. We serve the users in degraded conditions for improving QoS. The two algorithms proposed are implemented using EURANE Extensions for NS2 \cite{7}. The top level simulation structure of UMTS implementation in NS2 is shown in Fig \ref{fig:1}.
The main functionality additions to NS-2 come in the form of the RLC modes (AM and UM), mac-d/-c/sh support for HS-DSCH, i.e. HSDPA. The following are the associated TCL objects for this functionality:
\begin{itemize}
\item	Mac/Umts-Mac-d and Mac-c/sh
\item	Mac/Hsdpa-Mac-hs
\item	UMTS/RLC/AM-Normal AM
\item	UMTS/RLC/AMHS-Enhanced AM with flow/priority Packet status information.
\end{itemize}
The overall model of HSDPA protocol architecture is shown in Fig \ref{fig:2}.
The EURANE contributes three layers in NS-2. The first layer is RLC. It exists at the RNC and UE. The second layer is MAC which splits into two sub layers as mac-d which exits at RNC, UE and mac-hs which exists at node B (BS). The physical layer (PHY) exists at node B and UE. For the physical layer implementation, the standard NS-2 channel object is used to connect BS to UE's. The channel uses the standard NS-2 error model which provides a simple probabilistic packet loss. However, for HSDPA a more realistic error model is required to take into account the long- and short-term variations in link quality in order to model the fast link adaptation, fast scheduling and Hybrid ARQ (HARQ). 
\par
Therefore, EURANE uses a pre-computed Block Error Rate (BLER) /SNR performance curve and input trace file of received powers (expressed in SNR) to generate the required error model for HSDPA. The input trace file is generated for each UE according to the physical channel model. This file contains SNR values of the first, second and third transmission attempts in addition to the CQI in each time slot (2ms) \cite{13}. The files are then attached to the corresponding flow in NS-2 using the simulation script. The Mac/Hsdpa object in the Node B loads all the files into a memory data structure before the simulation starts. After that, the SNR and CQI values are read on demand by different class methods such as the scheduler and HARQ. For the SNR/BLER curve, a single file provides mapping between a certain BLER range and the minimum SNR value required for a specific CQI value. 
\par
BLER range and the minimum SNR value are required for estimating CQI values for each user. This file is also loaded into memory to be used by the receivers HsdpaPhy object to determine whether an ACK or NACK should be generated for the received SNR \cite{14}. The scheduling strategy has significant impact on the performance of the mac-hs flow control, as the amount of transferred data is dependent on its resource allocation. The scheduling is done at the BS (Mac/Hsdpa). The algorithm in general followed for scheduling is given below: 
\begin{enumerate}[S1:]
\item The BS initially receives the packet from RLC by using Sendown function.
\item For each packet the flow id and priority are considered and the index value is calculated using the formula\\                                      $\mbox{index} = \mbox{priority}* \mbox{flow}\_\mbox{max}+ \mbox{flow}_{ID}$, where flow\_max is set to 20 i.e. EURANE supports 20 mobile users.
\item The Packets are enqued at the respective index values.
\item The HARQ process information is noted down and for each flow the CQI values are obtained using the formula \cite{7}
		\begin{displaymath}
	CQI = \left\{ \begin{array}{ll}
	0 & SNR\leq -16\\
	\frac{SNR}{1.02}+16.62& -16<SNR<14\\
	30 & 14\leq SNR
	\end{array} \right.
	\end{displaymath}
	The SNR values in the above equation are pre-computed values obtained in MATLAB\textsuperscript{\textregistered} for each UE.
	\item The Scheduling method is being called at this step which takes the CQI values and HARQ process information as inputs. The scheduler checks the queue for every 2ms of TTI.
	\item Once the scheduling is done, the mac-d PDU's of each flow are assembled to create mac-hs PDUs.
	\item After this, the updation of HARQ process is done depending on receiving ACK/NACK.
	\item Finally the mac-hs PDUs will be assigned to the users correctly by the HARQ process.
	\item Repeat the steps from S5-S8 for every HARQ process.
\end{enumerate}

\section{Simulation Results and Discussion}
In this section we discuss about the result obtained from the work. The two modified scheduling algorithms has been tested in different environments.
The parameters considered for the simulation are given Table \ref{tab:1}.
\begin{table}[bpht!]
\centering
\caption{Simulation Parameters for Prioritized C/I Scheduling}
\label{tab:1}
\begin{tabular}{|l|l|}
\hline
\textbf{Parameter}&\textbf{Value} \\ 
\hline
Simulation Topology&	$500\times 500$ \\ \hline
Simulation Time&	10-80 sec \\ \hline
Transmission Time Interval(TTI)& 2ms \\ \hline
Channel Used&	HS-DSCH\\ \hline
Trace Length&	100sec\\ \hline
Number of Mobile Users &	10 \\ \hline
Distance from BS&	500m\\ \hline
Data Transfer Mode Used&	RLC AM\\ \hline
Traffic Type&	Exponential \\ \hline
\end{tabular}
\end{table}
The link characteristics (bandwidth and delay) are summarized in Table \ref{tab:2}. Only performances of radio interface between the BS and mobile users are examined. Mobile users are considered depending on their environment type.
\begin{table}[bpht!]
\centering
\caption{Link Characteristics}
\label{tab:2}
\begin{tabular}{|l|l|l|}
\hline
\textbf{Link}&\textbf{Bandwidth(Mbps)}&\textbf{Delay(ms)} \\ \hline
node 2-node 1&	100&	35\\ \hline
node1-ggsn&	100&	15\\ \hline
ggsn-sgsn	&622&	10\\ \hline
sgsn-rnc&	622&	0.4\\ \hline
rnc-bs(uplink)&	622&	15\\ \hline
rnc-bs(downlink)&	622&	15\\ \hline
\end{tabular}
\end{table}

Simulation results were obtained by tracing all packets through a wireless link from the BS to different mobile users. Data was collected from the trace file. Using the awk scripts the end to end delay is calculated. We compare the Prioritized C/I scheduling with Max C/I scheduling in various environments namely Pedestrian Environment, Rural Environment, Indoor Environment, Urban Environment and also on Hilly Environment.
\subsection{Pedestrian Environment}
In this environment we consider all the mobile users are moving at a speed of 3km/hr. As the Prioritized C/I scheduling takes priority into consideration instead of CQI values, unlike in the case of max C/I. This algorithm serves the user with highest priority. Here we consider the users who are requesting for different type of services and based on their priorities the scheduling is done. The comparison study reveals that by using this kind of scheduling the end to end delay for the user is reduced as compared to Max C/I based scheduling. 
\begin{figure}[bpht!]
\includegraphics[height=7cm,width=10cm]{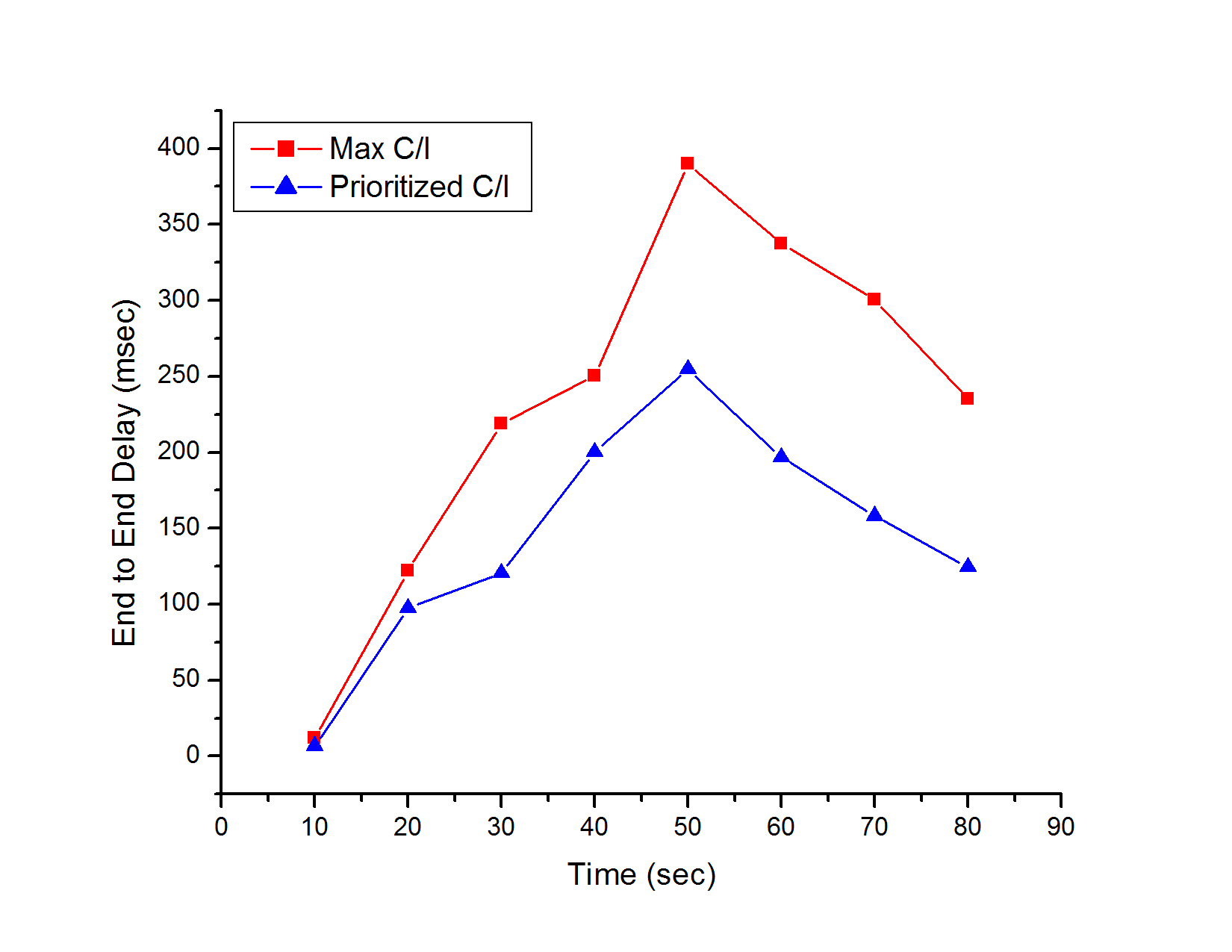}
\caption{Delay Performance of Prioritized C/I Scheduling Algorithm for Pedestrian Environment}
\label{fig:3}
\end{figure}
From the Fig.\ref{fig:3} we can observe the decrement after 50 sec time interval. This is because the numbers of retransmission are less. Hence we can say that packet delivery ratio has also improved with Prioritized C/I scheduling.
\subsection{Rural Environment}
In this environment we consider all the users are moving at a speed of 50 km/hr. 
\begin{figure}[bpht!]
\includegraphics[height=7cm,width=10cm]{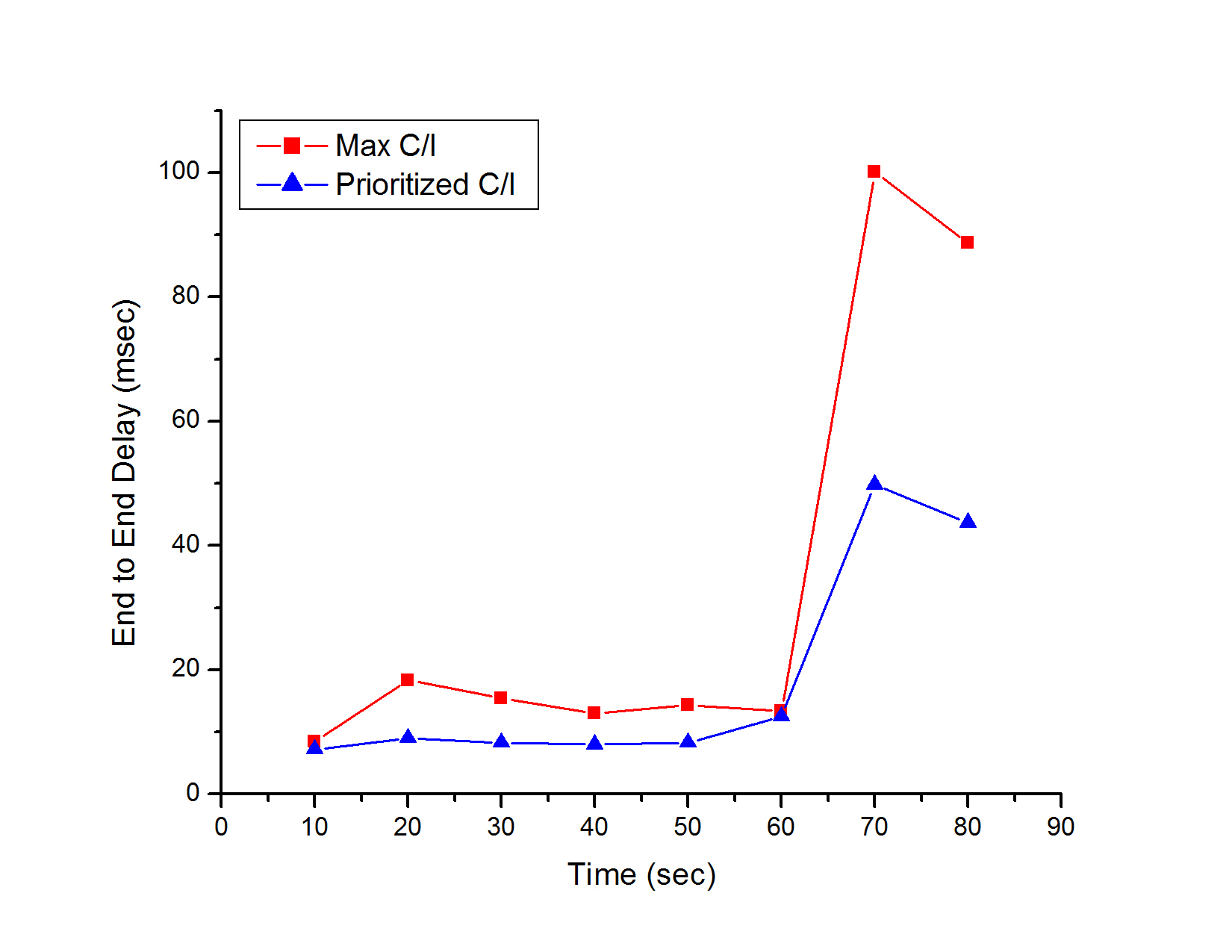}
\caption{Delay Performance of Prioritized C/I Scheduling Algorithm for Rural Environment}
\label{fig:4}
\end{figure}
As the speed of the mobile user increased the end to end delay observed when compared to Pedestrian Environment has decreased in both the scheduling algorithms for shorter distance from BS. In this environment also the performance of the prioritized C/I scheduling is better than Max C/I scheduling. The end-to-end delay performance for this environment is shown in Fig.\ref{fig:4}.
\subsection{Hilly Environment}
In the hilly environment mobile users are considered to be moving at a speed of 90 km/hr. The user speed has significant effect on the data transfer performances only when the distance from BS is less. The performance degrades when the distance from BS increases even when speed at which mobile user is increased. Hence here we choose an optimum distance from BS. 
\begin{figure}[bpht!]
\includegraphics[height=7cm,width=10cm]{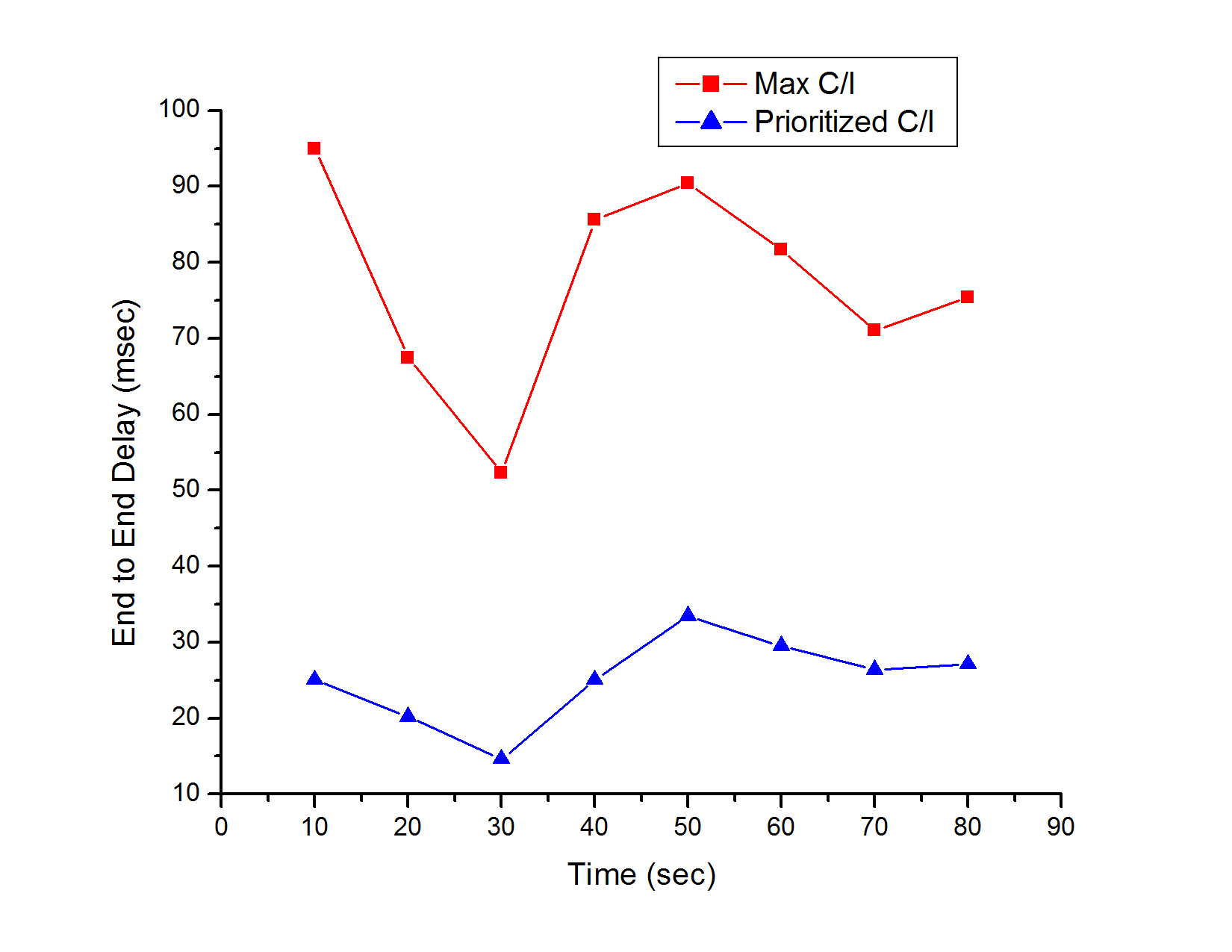}
\caption{Delay Performance of Prioritized C/I Scheduling Algorithm for Hilly Environment}
\label{fig:5}
\end{figure}
The end-to-end delay performance for this environment is shown in Fig.\ref{fig:5}.

\subsection{Indoor Environment}
\begin{figure}[bpht!]
\includegraphics[height=7cm,width=10cm]{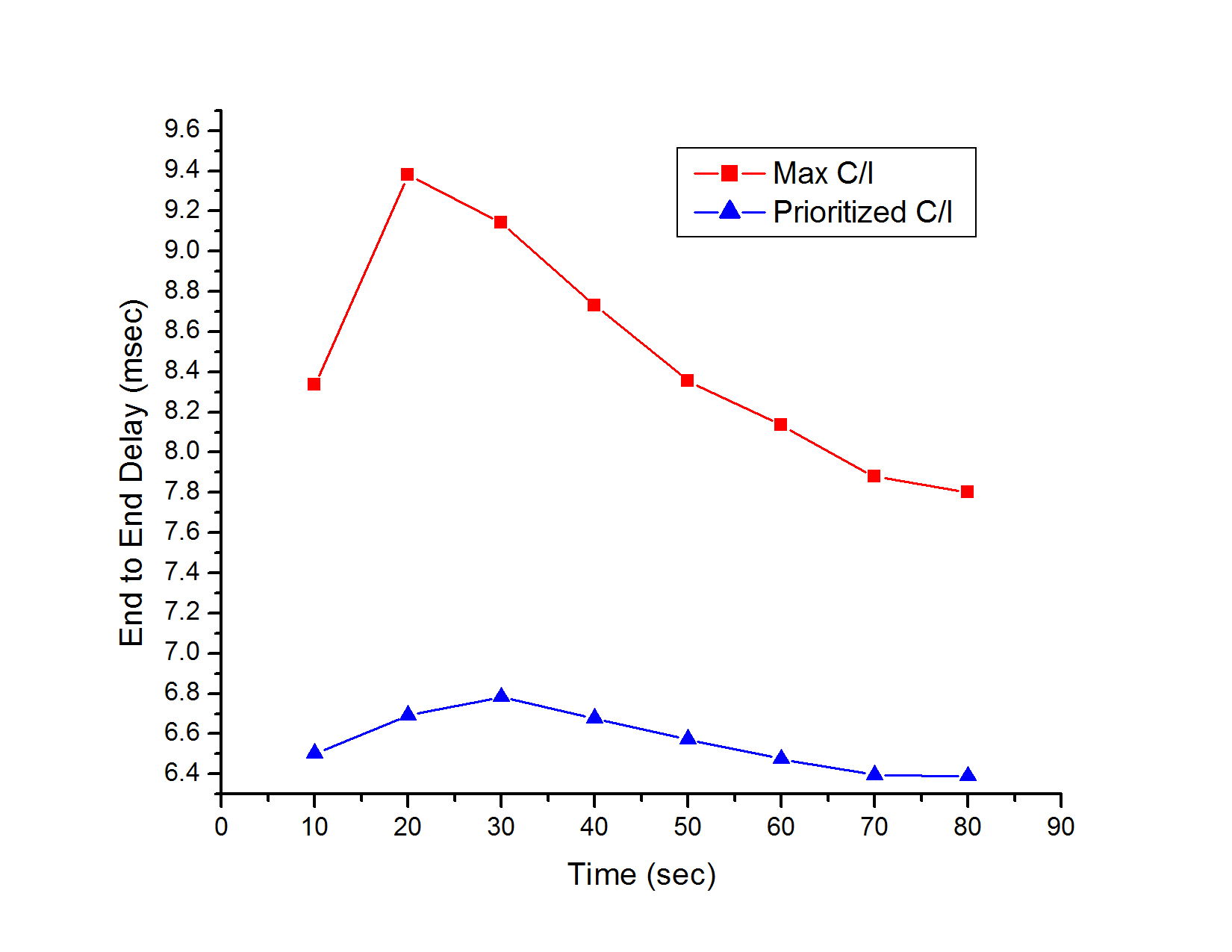}
\caption{Delay Performance of Prioritized C/I Scheduling Algorithm for Indoor Environment}
\label{fig:6}
\end{figure}

In the indoor environment mobile users are considered to be moving with the speed of 4km/hr. The graph is shown in Fig. \ref{fig:6}. From Fig.\ref{fig:6} we can infer that the smallest end to end delay is observed for the users who are present in this environment as they move with lesser speed. Hence we can conclude that the performance in Indoor environment is good.

\subsection{Urban Environment}
The mobile users are considered to be in urban environment moving at a speed of 100km/hr. Hence from the results obtained for five environments we can conclude that the performance of Prioritized C/I scheduling is found to be better when compared to Max C/I scheduling. Also, the mobile users speed has played a significant role in the analyzing the performance. When the speed of the users is less, and when all users are at equidistance from BS, as a consequence of the scheduling mechanism, the performance is better in terms of delay, thereby improving QoS.
\begin{figure}[bpht!]
\includegraphics[height=7cm,width=10cm]{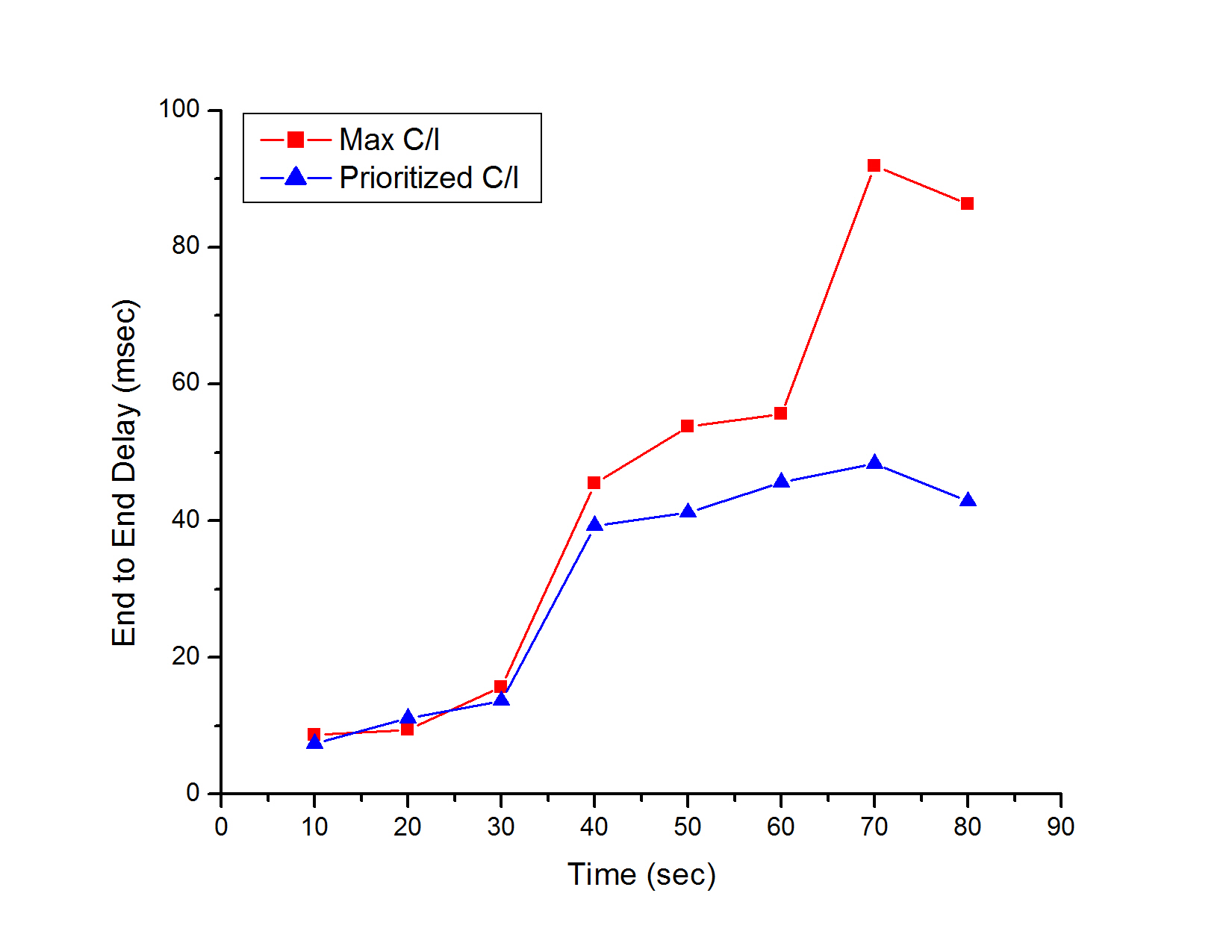}
\caption{Delay Performance of Prioritized C/I Scheduling Algorithm for Urban Environment}
\label{fig:7}
\end{figure}

\subsection{Performance of Modified Inverse C/I Scheduling}
\begin{figure}[bpht!]
\includegraphics[height=7cm, width=10cm]{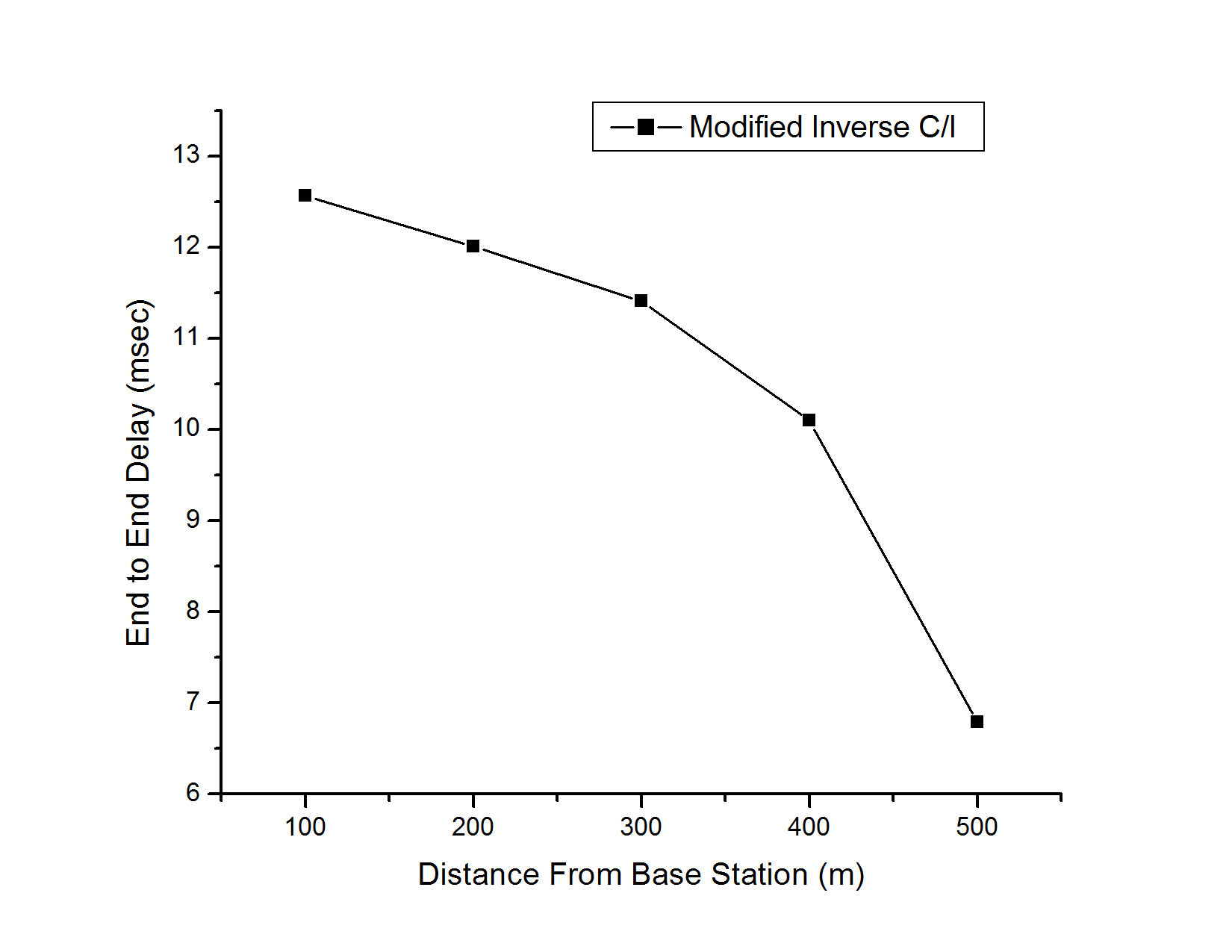}
\caption{Delay Performance of the Modified Inverse C/I Scheduling Algorithm}
\label{fig:8}
\end{figure}

We also modified another scheduling algorithm viz Inverse C/I based scheduling, for improving the quality of service for the users in degraded conditions.
The simulation parameter settings are given in Table \ref{tab:3}.
\begin{table}[bpht!]
\centering
\caption{Simulation Parameters for the Modified Inverse C/I Scheduling Algorithm}
\label{tab:3}
\begin{tabular}{|l|l|}
\hline
\textbf{Parameter}&\textbf{Value}\\ \hline
Simulation Topography& 	$500\times 500$ \\ \hline
Distances from BS	&100,200,300,400,500 mts \\ \hline
Simulation Time&	10 sec\\ \hline
Traffic Type& Exponential\\ \hline
Channel Used&	HS-DSCH \\ \hline
Data Transfer Mode& RLC AM \\ \hline
TTI&2msec \\ \hline
\end{tabular}
\end{table}
In this scheduling algorithm, the priorities are not assigned to any of the users initially. The user with least CQI value is considered and the priorities are given according to least CQI value. The graph is shown in Fig \ref{fig:8} for the Modified Inverse C/I scheduling.
The graph is plotted by considering 5 mobile users who are at different distances from BS. From Fig.\ref{fig:8} we can observed that with the Modified Inverse C/I scheduling the farthest mobile user with worst CQI value has highest priority as compared to the users with good CQI value hence improving the QoS for mobile users in the degraded condition.
\section{Conclusion}
Quality of Service (QoS) is an important issue as more number of multimedia services and more interconnections are increasing day by day. Hence there should be some techniques to be followed in improving the service quality. As the data services are increasing the service providers are facing difficulty in allocating the bandwidth for particular request. So the mobile users who pay less are getting degraded quality. In the network the amount of people who are willing to pay less will be more as compared to the users who pay more for getting good quality. So the user in the network should not leave the connection and switch to other network service provider which causes revenue loss to the original network service provider. Hence there should be better QoS for the users in degraded conditions.\par
 This paper concentrates on how the accessing delay is improved for users in different environments namely Pedestrian, Rural, Urban, Hilly and Indoor etc., requiring different services with varying priorities. We have proposed and implemented the Prioritized C/I scheduling mechanism for improving the end to end delay for the users who are equidistance from BS but requesting for different services. We also concentrated on the users who are having worst channel conditions, and implemented Modified Inverse C/I scheduling mechanism for improving the QoS for the users who are in degraded conditions.
Based on the simulation results it can be concluded that the performance is found to be better in terms of accessing delay with Prioritized C/I scheduling when compared to Max C/I scheduling. Also the Modified Inverse C/I has proved to give better QoS to the user in degraded condition hence improving the overall QoS.
\par
The present study is carried out in single cell environment by considering all the environments. The future work can include the multi cell environment with handoff possibilities. The performance analysis can be done by considering RLC in UM mode when the reliability is not given much of importance. The scheduling algorithms can also be tested in the upcoming LTE (3.9 G) technology.

\bibliographystyle{IEEEtran}
\bibliography{ref}

\end{document}